\documentclass[12pt]{article}
\usepackage{a4wide,epsfig}
\begin{document}    

\title{\vskip-3cm{\baselineskip14pt
\centerline{\normalsize\hfill TTP02--26}
\centerline{\normalsize\hfill hep-ph/0209357}
}
\vskip.7cm
The Impact of $\sigma(e^+e^-\to \mbox{hadrons})$ Measurements at Intermediate
Energies on the Parameters of the Standard Model
}

\author{
{J.H. K\"uhn}$^{a}$
and
{M. Steinhauser}$^b$
  \\[.7em]
  {\normalsize (a) Institut f\"ur Theoretische Teilchenphysik,}\\
  {\normalsize Universit\"at Karlsruhe, D-76128 Karlsruhe, Germany}
  \\[.5em]
  {\normalsize (b) II. Institut f\"ur Theoretische Physik,}\\ 
  {\normalsize Universit\"at Hamburg, D-22761 Hamburg, Germany}
}
\date{}
\maketitle

\begin{abstract}
\noindent
We discuss the impact of precision measurements of $\sigma(e^+e^-\to
\mbox{hadrons})$ 
in the center-of-mass range between 3 and 12~GeV, including improvements
in the electronic widths of the narrow charmonium and bottonium resonances,
on the determination of parameters of the Standard Model. 
In particular we discuss the impact of potential improvements on the 
extraction of the strong coupling constant $\alpha_s$, on the evaluation of 
the hadronic contributions to the electromagnetic coupling $\alpha(M_Z)$, and
the determination of the charm and bottom quark masses.
\end{abstract}


\section{Introduction}

In view of the possibility for improved measurements of the total cross
section in the energy region from approximately 3~GeV up to 12~GeV at
CLEO~\cite{CLEOc} it seems useful to analyze the relevance of measurements at
the different energy points for a variety of precision studies of the Standard
Model. The issues discussed in this brief note are:
\begin{itemize}
\item[(i)]
  the determination of the strong coupling $\alpha_s$,
\item[(ii)]
  the contributions from this region to the running of the fine
  structure constant,
\item[(iii)]
  the determination of the charm and bottom quark masses.
\end{itemize}
We will not be concerned with the interpretation of the (narrow and
wide) resonances in the context of quarkonium spectroscopy.

For definiteness we shall distinguish the following energy regions 
accessible by CLEO and
the corresponding contributions to the parameters of interest:
\begin{itemize}
\item[(R1)]
  The continuum below charm threshold $\langle 3~\mbox{GeV}, 2M_D\rangle$,
  excluding the narrow $J/\Psi$ and $\Psi^\prime$ resonances,
\item[(R2)]
  the charm threshold region $\langle 2M_D, 5~\mbox{GeV}\rangle$ with its
  rapidly varying cross section and wide charmonium resonances,
\item[(R3)]
  the continuum region below the bottom threshold 
  $\langle 5~\mbox{GeV},  2 M_B\rangle$, again excluding the narrow bottonium
  resonances $\Upsilon (1S)$, $\Upsilon (2S)$, $\Upsilon (3S)$,
\item[(R4)]
  the bottom threshold region $\langle 2 M_B, 11.5~\mbox{GeV} \rangle$ with
  its rapidly varying cross section and wide resonances,
\item[(R5)]
  the continuum region starting at 11.5~GeV,
\item[(R6)]
  the electronic widths of the narrow resonances $J/\Psi$ and $\Psi^\prime$,
\item[(R7)]
  the electronic widths of the narrow $\Upsilon$ resonances.
\end{itemize}
The separation points 5~GeV and 11.5~GeV should only be considered as
approximate and are chosen such that pQCD is valid at and above these
energies, an assumption to be tested by experiment.


\section{$\alpha_s$ and the validity of perturbative QCD}

\begin{figure}[t]
  \begin{center}
    \begin{tabular}{c}
      \leavevmode
      \epsfxsize=14cm
      \epsffile[40 220 550 550]{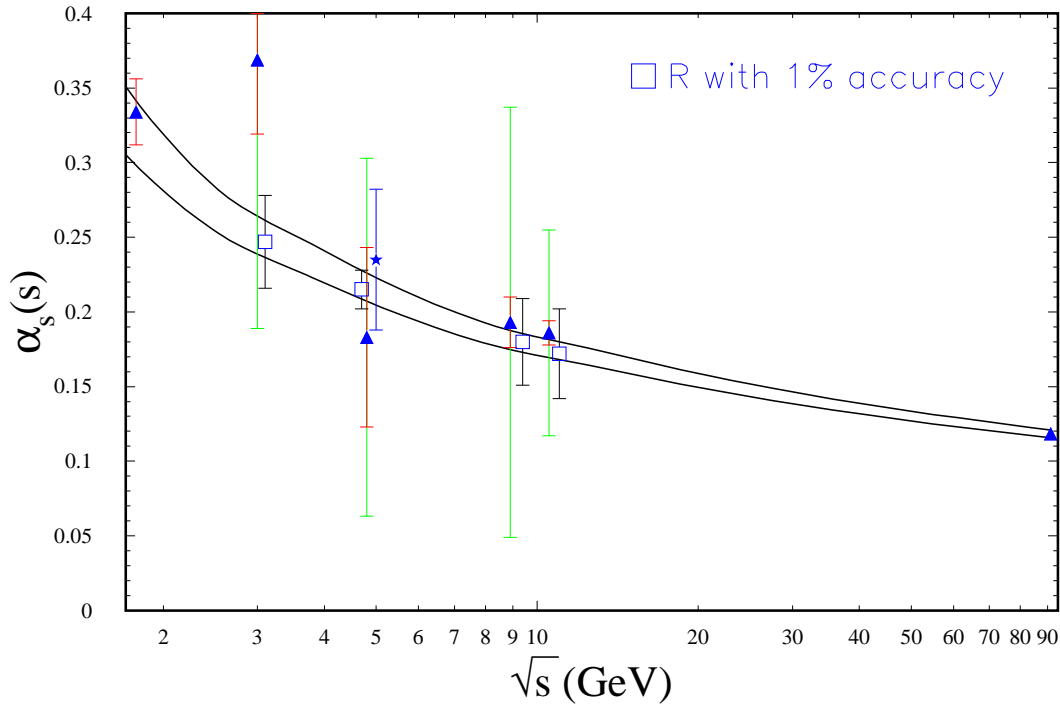}
    \end{tabular}
  \end{center}
  \vspace{-2.em}
  \caption{\label{fig::asrun}$\alpha_s$ as a function of $\sqrt{s}$.
    Results from $\tau$~\cite{ALEPHtau} and
    $Z$ decays~\cite{Gur00} and those extracted in~\cite{Kuhn:2001dm} from the
    $R$-ratio at 3~GeV, 4.8~GeV, 8.9~GeV and 10.52~GeV are shown.
    The two error bars on the data points indicate the statistical
    (inner) and systematical (outer) uncertainty.
    For the combined result (indicated by a star) 
    at $\sqrt{s}=5.0$~GeV only the 
    error after adding the statistical and systematical uncertainty in
    quadrature is shown.
    For illustration $\delta\alpha_s$
    reduced to the error as expected from $\delta R/R=1\%$ are also shown
    (slightly displaced in order to make the presentation more visible).
          }
\end{figure}

Predictions for $R(s)\equiv\sigma(e^+e^-\to \mbox{hadrons})/\sigma_{\rm pt}$ 
based on pQCD are valid down to fairly low energies. At present the agreement
between theory and experiment has been tested at the level of $2-4\%$ in the
energy region between 3 and 10.5~GeV. This has led to a determination of
$\alpha_s$ which already demonstrates the running of $\alpha_s$ as extracted
from the same observables, albeit at vastly different energies. The results as
derived from present experiments (including those derived from
$\tau$ and
$Z$ decays) are displayed in Fig.~\ref{fig::asrun}.
Measurements with precisions of 1\%, would be nearly competitive
with the determination of $\alpha_s$ from $\tau$ decays~\cite{ALEPHtau}
($\alpha_s(m_\tau)=0.334\pm0.022$) and
the hadronic $Z$-decay rate~\cite{Gur00} ($\alpha_s(M_Z)=0.1183\pm0.0027$)
and would lead to a beautiful confirmation of its running from $M_Z$ down
to $m_\tau$ as is demonstrated in Fig.~\ref{fig::asrun}.

\section{The continuum region and its relevance for the electromagnetic
  coupling}

Detailed predictions based on pQCD are available for the continuum
regions (R1), (R3) and (R5) 
(see~\cite{Chetyrkin:1996ia,Chetyrkin:1997pn,KHS} and references cited
therein).
Given $\alpha_s$, the
remaining uncertainty from unknown higher orders has been estimated to
be around 2.5\%, 1.5\% and  2.5\% for 3~GeV, 5~GeV and 11.5~GeV, respectively.
The validity of pQCD at these
points is taken for granted in all sum rule calculations (see below
Section~\ref{sec::moments}) 
and in the recent analyses of
$\alpha(M_Z)$~\cite{Chetyrkin:1997pn,Kuhn:1998ze,Davier:1998si} 
whereas the earlier papers employed pQCD only above 
40~GeV (see, e.g.~\cite{Eidelman:1995ny}).

\begin{table}[t]
  \begin{center}
    \begin{tabular}{l|c|c|c}
      energy region
      & (R1) & (R3) & (R5) \\
      \hline
      $\Delta\alpha^{(5)}_{\rm had}(M_Z^2)$ & $7.03\pm0.07$ & $41.72\pm 0.32$
      & $123.14\pm0.24$
    \end{tabular}
    \caption{\label{tab::delal}Contributions 
      to $\Delta\alpha^{(5)}_{\rm had}(M_Z^2)$ 
      (in units of $10^{-4}$) from the energy
      regions (R1), (R3) and (R5).}
  \end{center}
\end{table}
    
The absolute contributions to $\Delta\alpha^{(5)}_{\rm had}$
from the regions (R1), (R3) and (R5) 
(up to 40~GeV) based on pQCD are listed in Tab.~\ref{tab::delal}.
This has to be compared with a total contribution 
$\Delta\alpha^{(5)}_{\rm had}= 277.5 \pm 1.7 \times
10^{-4}$~\cite{Kuhn:1998ze}.
However, as emphasized above, this is based on the (though well founded)
{\em assumption} that pQCD is valid in this range and should be contrasted
with the present experimental uncertainties of 
4.3\% ((R1), BES~\cite{BES}), 4\% ((R3),MD1~\cite{MD-1})
and 2\% (10.52~GeV, CLEO~\cite{CLEO}).  
A measurement  of $\sigma(e^+e^-\to \mbox{hadrons})$ at a few well
chosen energy points would confirm or disprove the validity of pQCD 
in these regions and would allow to completely replace the theory driven
analysis by precise experiments combined with interpolations based on pQCD
or give additional support to the theory driven
evaluations based on pQCD.


\section{The impact of narrow resonances and the threshold region
  on $\Delta\alpha^{(5)}_{\rm had}$}

Narrow resonances contribute to $\Delta\alpha^{(5)}_{\rm had}$ through
\begin{eqnarray}
\Delta\alpha^{(5)}_R(M_Z^2) &=& 
                   \frac{3}{\alpha}\left(\frac{\alpha}{\alpha(M_R^2)}\right)^2
                   \frac{M_Z^2}{M_R^2}
                   \frac{M_R\Gamma_{ee}}{M_Z^2-M_R^2}
\,.
\label{delalres}
\end{eqnarray}
The contribution from the $\Upsilon$ resonances is smaller than the one 
of charmonium resonances by approximately one order of magnitude, a
consequence of the smaller charge of bottom quarks, and their larger mass.

The contribution from the lowest three charmonium resonances
to $\Delta\alpha^{(5)}_{\rm had}$ and its  present error is sizeable,
$9.24\pm0.74\times 10^{-4}$ to be compared with $56.90\pm1.10\times 10^{-4}$
from the low energy region up to 1.8~GeV, and also in comparison to the
total error of $1.68\times 10^{-4}$~\cite{Kuhn:1998ze}.

The same holds true for the threshold region (R2). For 
$\Delta\alpha^{(5)}_{\rm had}$ a significant improvement has already been
achieved by the BES collaboration, with their systematic error of roughly
4\%.
Nevertheless it would be desirable to reduce this error by another
factor two. This would again allow to replace the theory driven
treatment of the data  as described in~\cite{Kuhn:1998ze}
by a purely experiment-based evaluation. 
The contribution from region (R4) is less important in this context, if one
assumes the validity of pQCD for $u, d, s$ and $c$ production.

Hence the reduction of the systematic errors in the electronic widths
of the charmonium resonances 
and in the charm threshold cross section to approximately 2\% would lead to a
significant reduction of the uncertainty in 
$\Delta\alpha^{(5)}_{\rm had}$. In this
context it would not be necessary to arrive at this precision for every
individual scan point: it is only the weighted integral which matters.


\section{\label{sec::moments}The 
  determination of charm and bottom quark masses through moments}

Let us, in the first part, concentrate on the determination of the
charm quark mass.

The approach used in~\cite{Kuhn:2001dm} to compute the charm (and bottom) quark
mass is based on the use of low-order moments. This has the advantage that
non-perturbative effects from the gluon condensate can be neglected and no
resummation of the Coulomb singularities are required. As a consequence of
the latter one can directly determine the $\overline{\rm MS}$ quark mass
which is a big advantage as compared to those methods which intrinsically have
to deal with the pole mass.

On the theoretical side the computation of the moments is reduced to the
evaluation of the charm-quark contribution to the photon polarization
function for which the first eight terms are known analytically up to the
three-loop order~\cite{CKS}.

The quark mass can be extracted from the moments
\begin{eqnarray}
  {\cal M}_n^{\rm th} &=& 
  \frac{9}{4}Q_c^2
  \left(\frac{1}{4 m_c^2}\right)^n \bar{C}_n
  \,,
  \label{eq:Mth}
\end{eqnarray}
with coefficients $\bar{C}_n$ which depend logarithmically on $m_c$.
On the experimental side the moments can be spit into
three parts: 
\begin{eqnarray}
  {\cal M}_n^{\rm exp} &=& 
  \int \frac{{\rm d}s}{s^{n+1}} R_c(s)
  \nonumber\\&=&\mbox{}
  {\cal M}_n^{\rm exp,res}
  +{\cal M}_n^{\rm exp,cc}
  +{\cal M}_n^{\rm cont}
  \,,
\end{eqnarray}
the contribution from the resonances $J/\Psi$ and $\Psi^\prime$,
the contribution from the charm threshold region 
($3.73$~GeV$\le\sqrt{s}\le4.8$~GeV), and the  contribution from the continuum
above $\sqrt{s}=4.8$~GeV.
For ${\cal M}_n^{\rm exp,cc}$ the BES-data~\cite{BES} have been used.
Due to the use of low-order moments there is still a sizeable contribution
from ${\cal M}_n^{\rm cont}$. To be precise, it amounts to 31\% (10\%) for
$n=1$ ($n=2$).
A detailed decomposition of the individual contributions to the
moments and the error can be found in Tab.~\ref{tab:moments}.

\begin{table}[t]
\begin{center}
\begin{tabular}{c|c|c|c|c}
  & 
  \multicolumn{1}{c|}{$J/\Psi$, $\Psi^\prime$} & 
  \multicolumn{1}{c|}{charm threshold region} & 
  \multicolumn{1}{c|}{continuum} &
  sum \\
  \hline
  $n$ 
  & ${\cal M}_n^{\rm exp,res}$ 
  & ${\cal M}_n^{\rm exp,cc}$ 
  & ${\cal M}_n^{\rm cont}$ 
  & ${\cal M}_n^{\rm exp}$ \\
  & $\times 10^{(n-1)}$ 
  & $\times 10^{(n-1)}$ 
  & $\times 10^{(n-1)}$ 
  & $\times 10^{(n-1)}$
  \\
  \hline
$1$&$  0.1114(82)$ &$  0.0313(15)$ &$  0.0638(10)$ &$  0.2065(84)$ \\
$2$&$  0.1096(79)$ &$  0.0174(8)$  &$  0.0142(3)$  &$  0.1412(80)$ \\
\end{tabular}
\caption{\label{tab:moments}Experimental moments
  separated according to the contributions from
  the resonances, the charm threshold region and the continuum region
  above $\sqrt{s}=4.8$~GeV.}
\end{center}
\end{table}

Currently there is no reliable data for $R(s)$ in the energy region above 
$\sqrt{s}=4.8$~GeV. Thus, in~\cite{Kuhn:2001dm} for this part the theoretical
prediction for $R(s)$ has been used. 
This is motivated by the fact that there is very good agreement with
experiment in those energy regions where data is available
(see~\cite{Kuhn:2001dm}).
Note, that the full mass dependence for $R(s)$ is known up to order
$\alpha_s^2$, and the first three expansion terms 
in $m^2/s$ are available at order $\alpha_s^3$.
In~\cite{Kuhn:2001dm} the relative error of 
${\cal M}_n^{\rm cont}$ turned out to be 1.5\% (2\%) for $n=1$ ($n=2$).
However, as emphasized before, this consideration is based of the validity of
pQCD above 4.8~GeV. It would be of considerable importance to verify this
assumption through a precise measurement.

The charm quark mass cited in~\cite{Kuhn:2001dm} (obtained from $n=1$)
reads $m_c(3~\mbox{GeV})=1.027(30)$
which corresponds to $m_c(m_c)=1.304(27)$~GeV.
Almost $28$~MeV of the uncertainty in $m_c(3~\mbox{GeV})$ is of experimental
origin, i.e., comes from the error in ${\cal M}_n^{\rm exp}$.
In case the uncertainty in ${\cal M}_n^{\rm cont}$ is increased to 10\%
this increases to 35~MeV. 
Conversely, assuming that the error on the electronic widths of the narrow
resonances and the continuum could be reduced to 2\%, and adding the two
contributions with uncorrelated errors, the moments would be known with a
relative precision of roughly 1.5\%. This would lead to a final error on
$m_c(m_c)$ of 10 to 15~MeV.

To summarize:
a reliable measurement of $R(s)$ above the charm threshold region would be
very important to cross check or even replace the use of the theoretical
prediction for the evaluation of ${\cal M}_n^{\rm cont}$.
This would require a
scan of $R(s)$ for 4.8~GeV$\le\sqrt{s}\le$7.5~GeV.
Since the cross section is flat in this region a measurement, e.g., at three
or five different center-of-mass energies should be sufficient.
Furthermore, the uncertainties in $\Gamma_{ee}(J/\Psi)$ and
$\Gamma_{ee}(\Psi^\prime)$ and in the charm threshold region should be reduced
to 2\%. At the same time the separation of the $u,d,s$-background and the charm
contribution would be highly desirable. This would allow to test the
(plausible) assumption on the behaviour of the continuum cross section and
would furthermore lead to a reduction of the error in $m_c$ down to
$10-15$~MeV.

Similar considerations apply to the determination of $m_b$. In this case the
continuum (with the separation point presently chosen at
11.2~GeV~\cite{Kuhn:2001dm}) plays an even more important role 
(see Tab.~\ref{tab::momentsb} where the relative 
contribution to the lowest three moments from the three narrow resonances,
from the threshold contribution up to 11.2~GeV and from
the continuum above 11.2~GeV are listed).
Thus the measurement of $\sigma(e^+e^-\to b\bar{b})$ at
11.4~GeV, where a large data sample has already been collected would be
extremely useful for a test of pQCD at this energy point and thus for a
reliable evaluation of the continuum contribution to the low moments.
Given a 2\% measurement of $\Gamma_{ee}$ for the narrow resonances, a 2\%
measurement of the threshold region and a 2\% measurement at $11.4$~GeV, a
determination of $m_b(m_b)$ to 30~MeV seems well feasible.

\begin{table}[t]
\begin{center}
{
\begin{tabular}{l|lll|l}
$n$ & ${\cal M}_n^{\rm exp,res}$
& ${\cal M}_n^{\rm exp,thr}$
& ${\cal M}_n^{\rm cont}$
& ${\cal M}_n^{\rm exp}$
\\
  & $\times 10^{(2n+1)}$
& $\times 10^{(2n+1)}$
& $\times 10^{(2n+1)}$
& $\times 10^{(2n+1)}$
\\
\hline
$1$&$   1.237(63)$ &$   0.306(86)$ &$   2.913(21)$ &$   4.456(121)$ \\
$2$&$   1.312(65)$ &$   0.261(72)$ &$   1.182(12)$ &$   2.756(113)$ \\
$3$&$   1.399(68)$ &$   0.223(61)$ &$   0.634(8)$ &$   2.256(108)$ \\
\end{tabular}
}
\caption{\label{tab::momentsb}Moments for the bottom quark system:
  ${\cal M}_n^{\rm exp,res}$ includes the 
  contribution from $\Upsilon(1S)-\Upsilon(3S)$;
  ${\cal M}_n^{\rm exp,thr}$ includes the remaining 
  threshold contributions up to 11.2~GeV;
  ${\cal M}_n^{\rm cont}$ represents the continuum above 11.2~GeV.  
     }
\end{center}
\end{table}


\section{Summary}

The importance of the improved determination of the cross section for charm
and bottom production in their respective threshold region has been
discussed. Both regions are of importance for measurements of $\alpha_s$ in
the intermediate range. The charm region is of particular relevance for the
hadronic contribution to the electromagnetic coupling at $M_Z$. An improved
determination of $m_c$ and $m_b$ with a precision below 15~MeV and 30~MeV,
respectively, seems within reach, once these cross sections are known to
better than 2\%.
In general a precise determination of
$\sigma(e^+e^-\to \mbox{hadrons})$ at a few selected points, e.g. 3~GeV,
3.73~GeV, 5~GeV, 10.5~GeV and 11.5~GeV combined with a scan through the
threshold regions for charm and bottom production 
(and an evaluation of the weighted integrals) would be
sufficient for this purpose. The additional measurement of
$\sigma(e^+e^-\to \mbox{hadrons})$ at two or three selected points between 5
and 10~GeV could provide additional confidence in pQCD motivated
interpolations. 


\section*{Acknowledgement}
We thank D. Cassel for initiating this work and for discussions.


\end{document}